\documentstyle[12pt,epsfig,amssymb]{article}

\renewcommand
\baselinestretch{2}

\begin{document}
\setlength{\parskip}{0.45cm}
\setlength{\baselineskip}{0.75cm}
\begin{titlepage}

\vspace{1cm}
%
\vspace{0.5cm}
\begin{center}
\Large

{\bf
    Is the diffuse Gamma Background Radiation generated by
Galactic  Cosmic Rays?}\\
\vspace{0.5cm}
\large
 Arnon Dar$^{a,b}$ and A. De R\'ujula$^a$
\\

\vspace*{0.5cm}
\normalsize
$^a$ Theory Division, CERN, CH-1211 Geneva 23, Switzerland \\
$^b$ Technion, Israel Institute of Technology,
Haifa 32000, Israel\\
\vspace{0.7cm}
%
\large
{\bf Abstract} \\
\end{center}
\vspace*{-0.5cm}
\noindent
We explore the possibility that the diffuse gamma-ray background radiation
(GBR) at high galactic latitudes could be dominated by inverse Compton
scattering of cosmic ray (CR) electrons on the cosmic microwave background
radiation and on starlight from our own galaxy. 
Assuming that the
mechanisms accelerating galactic CR hadrons and electrons are the same, we
derive simple and successful relations between the spectral indices of the
GBR above a few MeV, and of the CR electrons and CR nuclei above a few
GeV. We reproduce the observed intensity and angular dependence of the
GBR, in directions away from the galactic disk and centre, without
recourse to hypothetical extragalactic sources.

\vspace{1.5cm}


\vspace{1.5cm}

\end{titlepage}
\newpage
\renewcommand
\baselinestretch{1}
\normalsize
\section{Introduction}

The existence
of an isotropic, diffuse gamma background radiation (GBR)
was first suggested by
data from the SAS 2 satellite (Thompson \& Fichtel 1982).
The EGRET instrument on the Compton Gamma Ray Observatory
confirmed this finding: by removal of point sources and of the
galactic-disk and galactic-centre emission, and after an extrapolation
to zero local column density,
a uniformly distributed GBR was found, of alleged
extragalactic origin (Sreekumar et al.~1998). Above an energy of  
$\sim\!10$ MeV, this radiation --to which we shall refer throughout simply
as ``the GBR''-- has a featureless spectrum, shown in Fig.~1,
which is very well described by a simple power-law form, 
${\rm dF/dE\propto E^{-\beta}}$, with $\beta\approx 2.10\pm0.03$ (Sreekumar et al.
1998). 

The origin of the GBR is still unknown.
The published candidate sources range from the
quite conventional to the decisively speculative.
Perhaps the most conservative hypothesis for the origin of an isotropic GBR
is that it is extragalactic, and originates from active galaxies (Bignami et 
al.~1979; Kazanas \& Protheroe 1983; Stecker \& Salamon 1996).
The fact that blazars have a $\gamma$-ray spectrum with an average index
$2.15\pm 0.04$, compatible with that of the GBR, supports
this hypothesis (Chiang \& Mukerjee 1998). 
The possibility has also been discussed
that Geminga-type pulsars, expelled into
the galactic halo by asymmetric supernova explosions,
be abundant enough to explain the GBR (Dixon et al. 1998;
Hartmann 1995). More exotic hypotheses include a baryon-symmetric
universe (Stecker et al.~1971), now excluded (Cohen et al.~1998), primordial black hole evaporation (Page \& Hawking 1976; Hawking 1977),  
supermassive black holes
formed at very high redshift (Gnedin \&  Ostriker 1992), annihilation of
weakly interactive  big-bang remnants (Silk \& Srednicki 1984; Rudaz  
\& Stecker 1991), and a long etc.
 
However, the EGRET GBR data in directions above the galactic disk and centre
show a significant deviation from isotropy, correlated with the
structure of our galaxy and our position relative to its centre
(Dar et al.~1999). This advocates  a local (as
opposed to cosmological) origin for the GBR.
Indications of a large galactic
contribution to the GBR at large latitudes were independently
found by Dixon et al. (1998) by means of a wavelet-based ``non-parametric''
approach that makes no reference to a particular model. 
Strong \& Moskalenko (1998) and Moskalenko \& Strong (2000)
also found that the contribution of inverse Compton scattering of galactic 
cosmic ray electrons to the diffuse $\gamma$-ray  
background is presumably much larger than previously thought.
In this paper we go one step further and explore in detail
the possibility (Dar et al.~1999)
that the diffuse gamma-ray background radiation
at high galactic latitudes could be dominated by inverse Compton
scattering of cosmic ray (CR) electrons on the cosmic microwave background
radiation and on starlight from our own galaxy. 
In Section  2 we briefly review    
the GBR data and the evidence for its correlation with our 
position in the Galaxy.

The CR-proton and CR-electron spectra
are briefly reviewed in Section 3.
The origin, spectrum and composition
of non-solar cosmic ray  protons and nuclei  have
been debated for almost a century.
The measurements now extend over
some 30 orders of magnitude in flux and some 15
orders of magnitude in energy, up to an astonishing
${\rm E}\!\sim\! 3\!\times\!10^{11}$ GeV
(Bird et al.~1995, Takeda et al.~1998, Berezinskii et al.~1990
and references therein). Above  
$\sim\! 5$ GeV,
this spectrum has also a power-law form ${\rm E^{-\beta}}$,
with two small
variations in the ``index'' $\beta$ at the so-called 
``CR knee''  and ``CR ankle''. The local spectrum 
of CR electrons, shown in Fig.~2,
is much harder to measure; it is only known up to
$\sim\! 10^3$ GeV and, above $\sim\! 5$ GeV, it is also well described
by a simple power law.

In Sections 4 and 5 we discuss relations between the indices
of the GBR and the CR electron and proton spectra. In so doing,
we make few and very simple
assumptions: that the mechanism accelerating CR hadrons and
CR electrons is the same (a moving magnetic ``mirror''),
that the locally-measured
electron spectrum is representative of its average form
throughout the Galaxy, that above a certain energy, inevitably,
the electron spectrum is modulated by inverse
Compton scattering on starlight and on the
microwave background radiation, and that the GBR
is dominated by the resulting Compton up-scattered
photons. This allows one to derive, successfully, the GBR index from
the electron index and the electron index from the
proton index.  The GBR index,
as observed by EGRET, is uncannily
directionally uniform. We interpret this fact as strong support
for our simple assumptions.

In Section 6 we tackle a more difficult and potentially
controversial subject: the origin and magnitude of the
GBR. In a sense, our proposed explanation
--that the GBR originates from inverse Compton scattering
in our own
galaxy (Dar et al.~1999) -- is more conservative than any of the 
previously suggested origins.

The non-conventional aspect of our hypothesis is that,
in order to reproduce the observed intensity of the GBR,
we must assume the scale height of our galaxy's
CR-electron distribution to be almost twice the traditionally-accepted
upper limit. Because of this, in Section 6, we briefly
review the basis of the conventional wisdom and our
critical view of it, whose main points are the following.
Moskalenko, Strong
and their collaborators have developed
a very detailed  understanding of the CR, radio and $\gamma$
observations of our galaxy. 
To fit the data, their models require a freely parametrized
reacceleration of electrons, presumably by the motion
of turbulent magnetic fields (e.g., Seo \& Ptuskin, 1994).
Strong \& Moskalenko  (1998) introduce a cutoff ${\rm z_h}$
for the height above the galactic plane above which
cosmic rays freely escape. They find an
upper limit ${\rm z_h\!<\!12}$ kpc,
on the basis of a fit to the ${\rm ^{10}Be/^9 Be}$ ratio observed by
Ulysses (Connell 1998). This result
is  ``soft'': twice the upper limit would still be compatible with
the ensemble of data (Lukasiak et al.~1994). Moreover,  the galactic
CR proton distribution extracted from a fit to EGRET $\gamma$-ray
data, actually favours (Strong \& Moskalenko 1998) an ad hoc distribution of
CR sources that is not as well localized in the disk as the
conventional supernova-remnant sources are (Webber 1997),
even if ${\rm z_h=20}$ kpc or more.
This point, and the necessity to invoke CR reacceleration,
indicate that scale heights of the CR electron distribution
in excess of the 12 kpc ``upper limit'' may not be out of the
question. Our results are optimized by a
scale height of roughly 20 kpc. 
Such a large scale height is not in contradiction
with radio synchrotron-emission from our galaxy if the galactic disk 
and its magnetic field are embedded  in a larger magnetic halo
with a much weaker field.

In studying the possibility that the diffuse GBR
is not extragalactic, one has two choices.
The first is to extend to high galactic latitudes
the elaborate models (with many parameters,
reacceleration, and ad hoc modifications
of the CR-proton and CR-electron energy and source distributions)
that have been developed to describe the intricate nature of the
observations at low galactic latitudes (Strong \& Moskalenko 1998;
Moskalenko \& Strong 2000).
The second is to adopt our very naive set of hypotheses
and employ a simple cosmic-ray model with, by conventional
standards, a large scale height for CR-electrons.
Models of this type (Dar \& Plaga 1999), wherein cosmic ray sources are
directly injected at high galactic
latitudes, have
actually been proposed\footnote{The injector agents
would be highly relativistic 
jets from the birth of compact objects in supernova explosions,
leading to a CR population permeating a magnetized region
of galactic-halo proportions and constituting a putative
solution to the problem of the origin of the highest-energy
cosmic rays,
a qualitative description of the nuclear CR spectrum, and a
possible explanation of jetted gamma-ray bursts.}.

In Section 7 we discuss the magnitude and angular-dependence
of the two dominant contributions to the GBR within our model: 
inverse Compton scattering of galactic CR-electrons
off the cosmic background radiation and starlight.
In Section 8 we compute the small additive effect
of sunlight, and in Section 9 we estimate the contribution
from external galaxies, which is also sub-dominant.
In Section 10 we compare our predictions with the data on the
intensity and the angular dependence of the GBR. The results
are very satisfactory and, within our model, lead to
the conclusion that the GBR can be dominated by the
emission from our own galaxy. We summarize our conclusions
and  predictions in  Section 11.

\section{The GBR data}
We call ``the GBR'' the
diffuse emission observed by EGRET
by masking the galactic plane at latitudes
$\rm{|b|\le 10^o}$, as well as the galactic centre
at $\rm{|b|\le 30^o}$ for longitudes $\rm{|l|\le 40^o}$,
and by extrapolating to zero column density, to eliminate the $\pi^0$
and bremsstrahlung contributions to the observed radiation and
to tame the model-dependence of the results.
Outside the mask, the GBR flux integrated over all directions in the 
observed energy range of ${\rm 30}$ MeV to ${\rm  120~GeV}$, shown in
Fig.~1, is well described by a power law:
\begin{equation}
{\rm {dF_\gamma\over dE}\simeq (2.74\pm 0.11)\times 10^{-3} \left
[E\over MeV \right]^{-2.10\pm 0.03}
~cm^{-2}~s^{-1}~sr^{-1}~MeV^{-1}}\; .
\label{photons}
\end{equation}
The overall magnitude
in Eq.~(\ref{photons}) is sensitive to the model used to subtract
the foreground (Sreekumar et al.~1998; Strong et al.~1998), but the  
spectral index is not.
The EGRET data are given in Sreekumar et al.~(1998) for 36 $\rm{(b,l)}$ domains, 9 values for each half-hemisphere.
The spectral index is, within errors, extremely directionally uniform,
as shown in Fig.~3, where we have plotted the EGRET results
as functions of $\theta$, the observation angle
relative to the direction to the galactic centre 
(${\rm \cos\theta=\cos\,[b]\, \cos\,[l]}$).
The normalization 
is less homogeneous, but in directions well above the galactic disk and 
away from the galactic-centre region it
has been found to be consistent with
a normal distribution around the mean value:
thus the claim of a possible extragalactic origin (Sreekumar et al. 1998).

In Fig.~4 we have plotted,  as a function of $\theta$, the EGRET GBR counting-rate above 100 MeV. This figure clearly shows,
in three out of the four quarters of the celestial sphere, an 
increase of the counting rate towards the galactic
centre. How significant is this effect? Let ${\rm \bar\chi^2\equiv\chi^2/d.o.f.}$
be the ``reduced'' $\chi^2$ per degree of freedom.
The $\bar\chi^2$ value for constant flux
is 2.6: very unsatisfactory. A best fit of the form
${\rm F=F_0+F_1\,(1-\cos\theta)}$  yields 
 ${\bar\chi^2=1.3}$, a very large amelioration (for higher
polynomials in $\cos\theta$ the higher-order coefficients are
compatible with zero: the fit does not significantly improve).
Note also that at angles with $\cos\theta$ larger than its mean value
$\langle\cos\theta\rangle\! =\! 0.0246$  (${\rm \theta<88.6^o}$), 10 out of the 12
 data points are above the average  flux, while at
angles with ${\rm \theta>88.6^o}$, 18 out of the 24  data points
are below the average. The probability for a
uniform distribution to produce this large or larger a
fluctuation is $1.5\times 10^{-4}$.

Even in directions pointing to the galactic disk 
and the galactic centre, 
EGRET data on $\gamma$-rays above 1 GeV show an excess over 
the expectation
from galactic cosmic-ray production of $\pi^0$'s (Pohl \& Esposito 
1998). Electron bremsstrahlung in gas is not the source
of the 1--30 MeV inner-Galaxy $\gamma$-rays observed by
COMPTEL (Strong et al.~1997), since their galactic latitude distribution is
broader than that of the gas. These findings also imply that inverse Compton scattering 
may be much more important than previously believed (Strong \& Moskalenko 1998; Moskalenko and Strong, 2000;
Dar et al.~1999).

\section{The CR data}

The cosmic ray  nuclei have a power-law spectral flux ${\rm
dF/dE\propto E^{-\beta}}$ with an index $\beta$ that changes at two
break-point energies. In the interval ${\rm 10^{10}~eV < E <
E_{knee}}$ $\sim 3 \times 10^{15}$ eV, protons constitute $\sim 96\%$
of the CRs at fixed energy per nucleon, and their flux is
(Berezinskii et al.~1990, and references therein):
 \begin{equation}
{\rm {dF_p\over dE}\simeq 1.8 \left
[E \over GeV \right]^{-2.70\pm 0.05}
~cm^{-2}~s^{-1}~sr^{-1}~GeV^{-1}}.
\label{protons}
\end{equation}
In the interval
$~{\rm E_{knee} < E < E_{ankle}}$ $\sim 3 \times 10^{18}$ eV,
the spectrum steepens from $\beta_1\sim 2.7$
to $\beta_2\sim 3.0$, flattening again to $\beta_3\sim 2.5$
above $\rm{ E_{ankle}}$.

The CR flux of electrons (Prince 1979; Nishimura et al.~1980; Tang 1984; Golden et al.~1984;  
Evenson \&
Meyers 1984; Golden et al.~1994; Ferrando et al.~1996; Barwick et al.~1998;
Wiebel-Sooth \& Biermann 1998), shown in Fig.~2,
is well fitted, from  ${\rm E\sim 10~GeV}$  to $\sim 2$ TeV  by:
\begin{equation}
{\rm {dF_e\over dE}\simeq (2.5\pm 0.5)\times 10^5 \left
[E \over MeV \right]^{-3.2\pm 0.10}
~cm^{-2}~s^{-1}~sr^{-1}~MeV^{-1}}.
\label{electrons}
\end{equation}
The terrestrial and solar magnetic fields and the solar wind modify
the electron
spectrum below  ${\rm E\sim 10}$ GeV, so that
the direct observations at those energies
 may deviate from the local interstellar spectral shape.

Cosmic ray electrons undergo inverse Compton scattering
(ICS) off the ambient photon baths:
starlight and the cosmic background radiation.
The spectral indices of the GBR and
electron spectra 
can be very simply and successfully related (Dar et al.~1999), if
the GBR dominantly consists of photons whose energy has been
uplifted by ICS, as we proceed to show.

\section{The index of the GBR spectrum}

The current temperature, number density and mean
energy of the CMB are ${\rm
T_0=2.728}$
K, ${\rm n_0\approx 411~cm^{-3}}$, and
${\rm \epsilon_0\approx} $ ${\rm 2.7\, kT_0\approx }$ ${\rm  
6.36\times10^{-10}~ MeV}$ (Mather et al.~1993; Fixsen et al.~1996).
The galactic starlight (SL) distribution is highly non-uniform,
its average energy is $\epsilon_\star\!\sim \!1$ eV.
Consider the ICS of high energy electrons on these radiations.
Assume the shape of the electron
flux, Eq.~(\ref{electrons}), observed at ${\rm E>10}$ GeV,
to be representative of the average galactic spectrum.
For the energy range of  EGRET 
the Thomson limit is accurate
even for ICS on SL, and the ${\rm e\gamma}$ cross section
is ${\rm \sigma_{_T}\approx 0.65\times 10^{-24}~cm^2}$.
The mean energy ${\rm E_\gamma}$ of the upscattered photons,
--or ${\rm \Delta E_e}$, the mean energy loss per collision-- is:
\begin{equation}
{\rm E_\gamma(\epsilon_i) \approx \Delta
E_e (\epsilon_i)\approx {4\over 3}\,\left({E_e\over m_e\,
c^2}\right)^2\,\epsilon_i} \, ,
\label{loss}
\end{equation}
with ${\rm \epsilon_i=\epsilon_0}$ or $\epsilon_\star$.

The ICS photon spectrum originating in our galaxy is the sum of CMB
and SL contributions:
\begin{equation}
{\rm {dF_\gamma\over dE}=
{dF_\gamma^0\over dE}+
{dF_\gamma^\star\over dE}}\, ,
\label{ICSphotons}
\end{equation}
and is a function of the galactic latitude (b) and longitude (l) coordinates.
The ICS final-photon spectrum --a cumbersome
convolution (Felten \& Morrison 1966) of a CR power spectrum with a
photon thermal distribution-- can be approximated very simply.
Using again the index ``i'' to label the CMB and SL fluxes:
\begin{equation}
{\rm {dF^i_\gamma\over dE_\gamma}
\simeq {N_i(b,l)~\sigma_{_T}}~{dE^i_e\over dE_\gamma}~
\left[ {dF_e\over dE_e}
\right]_{E_e=
{E}_e^i}\,;~~~~  {E}_e^i\equiv m_ec^2
\sqrt{{3\, E_\gamma\over 4\,\epsilon_i}}},
\label{ICSCBR}
\end{equation}
where ${\rm E_e^i}$ is obtained from Eqs.~(\ref{loss}) by inverting
${\rm E_\gamma(\epsilon_i)}$.
We postpone to Section 6 the discussion
of the model-dependent normalization factors
${\rm N_\star(b,l)}$ and ${\rm N_0(b,l)}$: effective  column densities resulting from the
convolution of the space distribution of CR electrons 
with those of starlight and of the CMB.
Introducing the CR-electron flux of 
Eq.~(\ref{electrons}), of the form ${\rm dF_e/dE=A\,[E/MeV]^{-\beta_e}}$, into Eqs.~(\ref{ICSCBR}), we obtain:
\begin{equation}
{\rm {dF^i_\gamma\over dE}=
{N_i(b,l)~\sigma_{_T}\,A\over 2}~
\left[{4\,\epsilon_i\,MeV\over 3\,m_e^2c^4}\right]^{{\beta_e-1\over 2}}
\,\left[{E\over MeV} \right]^{-{\beta_e+1\over 2}}
\propto
[E ]^{-2.10\pm 0.05}}\, .
\label{ICSphotpred}
\end{equation}
In the energy-range of EGRET, the CMB and
SL contributions have the same spectral index, 
as do the small sunlight and external-galaxy contributions 
discussed in Sections 8 and 9.

 The photon spectral index of  Eqs.~(\ref{ICSphotpred}),
which is related to that of the CR-electrons
through $\beta_\gamma=(\beta_e+1)/2$,
coincides with the measured one, Eq.~(\ref{photons}).
The electron spectrum of Eq.~(\ref{electrons})
describes the data in the range ${\rm E_e>5}$ GeV, so that
Eq.~(\ref{ICSphotpred}) should be valid above ${\rm E_\gamma\sim
100}$ keV, the typical energy of photons up-scattered from the CMB.
At ${\rm E_\gamma>
50}$ GeV, at the upper end of the EGRET data,
${\rm \sigma_T}$ in the SL contribution should be replaced
by the complete Klein--Nishina cross section, implying a steepening 
of the spectrum. The corresponding 
effect for the CMB contribution is at energy above the EGRET energy range.

In deriving Eqs.~(\ref{ICSphotpred}), we have assumed that
the locally-measured slope of Eq.~(\ref{electrons})
is representative of the index of the spectrum of the electrons
suffering ICS to produce the GBR, wherever they may be.
The spectral index of the diffuse GBR observed by EGRET is
independent of direction, as shown in Fig.~3.
The statistical test for a flat distribution is surprisingly good:
${\bar\chi^2\sim 0.5}$. This is encouraging support for our working
hypothesis of an electron spectrum with a universal shape, and of
a simple and dominant mechanism --ICS-- to generate the GBR.

\section{The index of the electron spectrum}

To relate the spectra of CR electrons and protons,
we need an estimate of the protons' spectrum at their source.
 A source spectrum
${\rm dF^s/dE}$ with index ${\rm\beta_s\sim 2.2}$
is obtained from collisionless shock simulations (Bednarz \& Ostrowski 1998)
or analytical estimates of acceleration by relativistic jets (Dar 1998).
The CR spectrum of nuclei is modulated by their
residence time in the Galaxy, ${\rm \tau_{gal}(E)}$. For a steady source of
CRs the energy dependence of the observed flux is roughly that of
${\rm \tau_{gal}\, dF^s/dE}$.
Observations of astrophysical and solar plasmas and of nuclear  
abundances as functions of energy (e.g. Swordy et al.~1990) indicate  
that
${\rm \tau_{gal}(E)\propto E^{ -0.5\pm 0.1}}$,
explaining ${\rm \beta_1\sim \beta_s +0.5 \sim 2.7}$, as in Eq.~(\ref{protons}).

Practically all CR acceleration mechanisms invoke
an ionized  medium
that is swept by a moving magnetic field, such as would
be carried by the rarefied plasma in a supernova shell (Bhattacharjee  
\& Sigl 2000)
or by a `plasmoid' of jetted ejecta (Dar \& Plaga 1999).
The magnetic field acts as a moving `mirror' that imparts the
same distribution in velocity, or Lorentz factor
${\rm \gamma=E/m\,c^2}$, to all charged
particles. To the extent that particle-specific losses
(such as synchrotron radiation) can be neglected
at the acceleration stage, all source fluxes
have the same energy-dependence. For electrons
below the anticipated `electron's knee' at
${\rm E_e=(m_e/m_p)\, E_{knee}}$$\sim 2$ TeV,
we expect ${\rm dF^s_e/dE\propto E^{-\beta_s}}$, with
$\beta_s\sim 2.2$.
Confinement effects preserve
this equality for ultrarelativistic electrons and protons: their
behaviour in a magnetic maze is the same.
But, unlike for hadrons, the `cooling' time of electrons
--that are significantly affected by the ambient radiation
and magnetic fields-- is shorter than their galactic
confinement time,
${\rm \tau_{gal}(E)}$, above a relatively low energy. This implies
that the CR electron spectrum is modulated mainly by the ICS, and not
by the confinement time.

Electrons lose energy not only by ICS on starlight and the CMB, but also  by
synchrotron radiation on magnetic fields. All of these processes are
essentially the same: scattering off photons, either real or virtual.
The energy loss is governed by the
rate at which a single electron interacts with the ambient  
electromagnetic fields, weighted by the corresponding average energy  
density:
${\rm P=\sigma_{_T}\, c\,[n_\star\epsilon_\star+n_0\epsilon_0+B^2/(8\pi)]}$.
Let ${\rm R_p}$ (an inverse time) be the production rate of CR  
electrons, assumed to be constant (Berezinskii et al.~1990), and let  
${\rm dn^s_e/dE}$ be their source
number-density spectrum. The actual density ${\rm dn_e/dE}$
in an interval ${\rm dE}$ about ${\rm E}$ is
continuously replenished and depleted
by electrons whose energy is being degraded by interactions. This leads to
a steady-state situation in which production and losses are in balance. Using Eq.~(\ref{loss}) we obtain:
\begin{equation}
{\rm {4\over 3}\,{P\over (m_e\, c^2)^2}
\, {d\over dE} \left(E^2\, {dn_e\over
dE}\right)=
      R_p\,{dn^s_e\over dE}}\, .
\label{balance}
\end{equation}

For a relatively uniform galactic CR
electron density, Eq.~(\ref{balance}) also applies to the
local electron flux ${\rm dF_e\simeq (c/4\pi)dn_e}$.
Substitute the spectrum ${\rm
dn^s_e/dE\sim E^{-\beta_s}}$
into the flux version of Eq.~(\ref{balance}) to
obtain:
\begin{equation}
{\rm {dF_e\over dE}= {3\, m_e^2\, c^4\, R\over
4\,(\beta_s-1)\,P}\; { dF^s_e\over
 E\, dE}\propto E^{-(\beta_s+1)}}\, .
\label{electron}
\end{equation}
For electrons with ${\rm E_e<(m_e/m_p)\, E_{knee}}$
we deduced that
$\beta_s\sim 2.2~.$ Thus, $\beta_s+1=3.2$,
in agreement with the data:
Eq.~(\ref{electrons}) and Fig.~2.
Above the `electron's knee' at ${\rm E_e\sim 2}$ TeV
the spectrum should steepen up by $\Delta
\beta \simeq 0.25$, like that of CR hadrons (Dar 1998). The available
spectral measurements extend only to ${\rm E_e\leq 1.5}$ TeV.

The energy density in the CMB is ${\rm n_0\epsilon_0=0.24}$ eV cm$^{-3}$,
coincidentally similar to that in starlight at our location:
${\rm n_\star\epsilon_\star\sim 0.22}$ eV cm$^{-3}$. If the local CR
and magnetic energy densities are in equipartition,
${\rm B^2/(8\pi)\sim 1}$  eV cm$^{-3}$, again in the same ballpark.
The cooling time of electrons in the ensemble of these fields is:
\begin{equation}
{\rm \tau_{_{cool}}(E)\simeq {3\,m_e^2\, c^4\over
 4 \, P\, E}
\simeq 0.22\times \left [{E\over GeV}\right]^{-1}~Gy}\, .
\label{coolingbis}
\end{equation}
The galactic escape time of GeV electrons, which should be similar to that of
CR protons
${\rm \tau_{gal}(E)\propto E^{ -0.5\pm 0.1}}$ (Swordy et al.~1990),
has a weaker energy dependence than that of  $\rm \tau_{_{cool}}$.
At sufficiently low energy, then, ${\rm \tau_{gal}<\tau_{_{cool}}}$,
and processes other than Compton- or synchrotron cooling
(such as Coulomb scattering, ionization losses and bremsstrahlung)
 become relevant.
The slope of Eq.~(\ref{electron})
should change as the energy is lowered. The spectrum of
Fig.~2 shows such a change, but it occurs at ${\rm E<10~GeV}$,
a range in which local modulations would mask the effect.

\section{The scale height of CR electrons}

The radio emission of galaxies seen edge-on --interpreted as
synchrotron radiation
by electrons on their local magnetic field--
offers direct observational evidence for CR electrons well above
galactic disks (e.g. Duric et al.~1998). For the particularly well observed
case of NGC 5755, the exponential
scale height of the synchrotron radiation is  ${\cal{O}} (4)$ kpc.
If the CRs  and the magnetic field energy are in equilibrium, they
should have
similar distributions, and the exponential
scale height ${\rm h_e}$ of the electrons ought to
be roughly twice that of
the synchrotron intensity, which reflects the convolution of
the electron- and magnetic-field distributions. The
inferred value ${\rm h_e}\sim 8$
kpc for NGC 5755 may not be universal for spirals, since ${\rm h_e}$ is very
sensitive to the density and distribution of CR sources, gas and plasma
in each particular galaxy. Moreover, the magnetic field may be in equipartition with cosmic rays only where the interstellar plasma is dense enough.  It is
quite possible for the CR electrons to be confined in a large
magnetic halo with a  field much smaller than that in the disk.
For these reasons we must discuss the observations
of our own particular galaxy.

Traditionally CR electrons and nuclei were assumed to have a distribution
that snugly fit that of the visible part of the Galaxy --where their
conventional sources lie-- implying a scale height above the plane of  
the disk
of ${\cal{O}} (1)$ kpc (Broadbend et al.~1989). As the data and their  
analysis
became more
elaborate, scale heights more than one order of magnitude larger were
discussed (e.g. Strong et al.~1998). Since electrons lose energy to the ambient
radiation close to their sources, which have traditionally been
located in the
disk, not very well understood
CR-reacceleration phenomena have had to be
invoked (e.g. Seo \& Ptuskin 1994).
Even with reacceleration, a conventional distribution of cosmic-ray
sources fails to describe the observed GBR (Strong \& Moskalenko 1998).

Over the years,
Moskalenko, Strong and their collaborators have developed
what is presumably the most
elaborate and detailed understanding of the CR, radio and $\gamma$
observations of our galaxy
(Moskalenko et al.~1998; Moskalenko and Strong, 2000;
Strong and Moskalenko, 1998; Strong et al.~1997;
Strong et al.~1998). A crucial parameter in their models is the scale ${\rm z_h}$
of the CR distribution orthogonal to the galactic plane,  defined as
the height above which CRs freely escape, as in a leaky-box model.
Strong \& Moskalenko (1998)  conclude
that   ${\rm z_h}$ lies between 4 and 12 kpc. The limits are based on the
comparison of the ${\rm ^{10}Be/^9Be}$ ratio observed by
Ulysses (Connell 1998)
with model predictions as a function of ${\rm z_h}$, being all other
parameters
fixed at their adopted values. The dependence of the ${\rm ^{10}Be/^9Be}$
ratio on  ${\rm z_h}$, shown in Fig.~9 of Strong \& Moskalenko (1998)
and reproduced here as Fig.~5, is very weak for
${\rm z_h>10}$ kpc. At ${\rm z_h=20}$ kpc, the prediction would
be only some  1.3 standard
deviations below the Ulysses central value, and even ${\rm z_h=40}$ would
be viable: the average of all previous and somewhat less precise
observations,
compiled in Lukasiak et al.~(1994) and shown in
Fig.~5a, would be in  agreement with ${\rm z_h=}$
20 or 40 kpc. For all these reasons and
the ones stated in the introduction, we shall not refrain from considering
scale heights above the 12 kpc upper limit quoted by Strong \&  
Moskalenko (1998).

\section{The CBB and SL contributions to the GBR}

The spectral index of the GBR,  derived in Section 4,
is independent of the details of the spatial distribution
of starlight. We have argued that the EGRET GBR data
support the simple hypothesis of an electron spectral
index that is independent of location. The predicted
GBR index is then also independent of the magnitude of the
electron spectrum as a function of position.
In this section we use a simplified model of the electron and starlight
distributions to compute the magnitude and angular
dependence of the CMB and SL contributions to the GBR.

We adopt ${\rm h_e=20}$ kpc (a value obtained from a
rough fit of our results to the angularly-averaged fluence of
the GBR) for the Gaussian
scale height of the CR electron distribution
of our galaxy in the direction perpendicular to the galactic plane.
For the
distribution in $\rho$ --the radial coordinate orthogonal to the
galactic axis-- we
adopt a Gaussian scale height $\rm{\rho_e=35}$ kpc; the results are
quite insensitive to this
parameter. The EGRET GBR data are not precise enough
to be ``invertible'', that is, for the
actual high-latitude CR-electron distribution (Gaussian, exponential
or otherwise) to be disentangled; a fact to be rediscussed anon,
in view of our results.
The distance of the solar system to the galactic centre
is ${\rm d_{_\odot}\simeq 8.5}$ kpc. The factor
${\rm N_0(\theta,\phi)}$ in Eq.~(\ref{ICSCBR}), which describes the  
angular dependence
of the GBR photons due to ICS on the (uniformly distributed) CMB, is:
\begin{eqnarray}
&&{\rm N_0(b,l)= \int_0^\infty \, dr \;n_0\;
Exp\left[+\left({d_{_\odot} \over \rho_e}\right)^2\right]
Exp\left[-\,\left({h(r,b)\over h_e}\right)^2
-\left({\rho(r,b,l)\over \rho_e}\right)^2\,\right]}\, ,\nonumber \\
&&{\rm h(r,b)\equiv r\,sin\,b }\, ,\nonumber \\
&&{\rm \rho(r,b,l)\equiv
\left([r\,\cos\,(b)\,\cos\,(l)-d_{_\odot}]^2+[r\,\cos\,(b)\,\sin\,(l)]^2\right)^{1/2}}\, ,
\label{CBRangle}
\end{eqnarray}
where r is the distance in the direction along the line of sight.

It is difficult to model in detail the contributionn from ICS on
starlight (Hunter et al.~1997, Sreekumar et al.~1998). But we are only  
concerned with this light
at high galactic latitudes, since the diffuse GBR of interest to us is that
measured by EGRET by masking the galactic plane and
centre.
We make a coarse estimate by approximating the Galaxy's starlight
as that produced by a source at its centre with the galactic
luminosity ${\rm L_\star}=2.3\times 10^{10}$ ${\rm L_{_\odot}}$
$\simeq 5.5~10^{55}$ eV s$^{-1}$ (Pritchet \& van den Bergh 1999).
The starlight
 contribution in Eq.~(\ref{ICSphotons}) is then of the same form as
Eq.~(\ref{CBRangle}),
with  ${\rm N_0}$ traded for ${\rm N_\star}$ by the substitution:
\begin{eqnarray}
{\rm n_0\rightarrow {L_\star\over 4\,\pi\,c\,\epsilon_\star}\;\;
{1\over( r^2-2\,r\,d_{_\odot}\,\cos\,(b)\,\cos\,(l)+d_{_\odot}^2)}}\, .
\label{Lightangle}
\end{eqnarray}

For the CMB and starlight contributions to the GBR,
averaged over the EGRET unmasked domain, we obtain, by integration of
Eqs.~(\ref{ICSphotons}), (\ref{ICSCBR}), (\ref{CBRangle}), (\ref{Lightangle}):
\begin{equation}
{\rm {dF_\gamma\over dE}
\simeq (2.41\pm 0.55)\times\! 10^{-3}\left[
 E\over MeV \right]^{-2.10\pm 0.05}
cm^{-2}s^{-1}sr^{-1}MeV^{-1}}.
\label{ICSphotons2}
\end{equation}
For  scale heights  ${\rm h_e}$ and $\rho_{\rm e}$ similar to
the ones adopted  (20 and 35 kpc,
respectively), the CMB and SL contributions are comparable
in magnitude, the first scales approximately linearly
with ${\rm h_e}$ while the second  is rather
insensitive to this parameter. The contribution to the CMB
from sunlight and external galaxies, discussed in
Section 8 and 9, adds corrections of 6\% and 
$\sim \! \!10\%$ (respectively) to Eq.~(\ref{ICSphotons2}),
the total result is shown in Fig.~2.
The fitted value of ${\rm h_e}$ is imprecise: the starlight to CMB ratio
is proportional to $\epsilon_\star/\epsilon_0$ raised to a
very poorly determined power, $0.10\pm 0.05$.

We can  use our assumed Gaussian distribution of
electrons in a  halo, with vertical and radial
scale heights ${\rm h_e}$ and ${\rm \rho_e}$, to compute 
the  diffuse $\gamma$-ray luminosity of our galaxy,
which in our model is dominated by ICS on CMB and SL photons.
Using 
Eqs.~(\ref{ICSphotons}), (\ref{ICSCBR}), (\ref{CBRangle}), (\ref{Lightangle})
we obtain, for the luminosity in
$\gamma$-rays of energy above E:
\begin{eqnarray}
&&{\rm L_\gamma (>E) \simeq L^0_\gamma (>E)
     +L^\star_\gamma(>E)}\, ,\nonumber \\
&&{\rm  L^0_\gamma (>E)=1.31\times 10^{40}
\left[{\rho_e\over 35~kpc}\right]^2
\left[{h_e\over 20~kpc}\right]
\left[{E\over MeV}\right]^{-0.10\pm 0.05}~erg/s} ,\nonumber \\
&&{\rm  L^\star_\gamma (>E)= 3.56\times 10^{39}
\left[{h_e\over 20~kpc}\right]\, \left[{1\over 2\, u}\, ln{1+u\over 
1-u}\right] \left[{E\over MeV}\right]^{-0.10\pm 0.05}~erg/s},
\label{radiation}
\end{eqnarray}
where ${\rm u\equiv \sqrt{1-h_e^2/\rho_e^2}}$.
A future $\gamma$-ray telescope, such as GLAST, could possibly
see the corresponding glow of Andromeda's halo.

\section{Sunlight contribution to the local GBR}

We are only at a distance ${\rm l_{_\odot}=1.5\times 10^{13}~cm}$
from the sun. This
entails a small but non-negligible
contribution to the locally-observed GBR, resulting from
ICS off photons in the heliosphere. The corresponding photon
flux is described by
Eq.~(\ref{ICSphotpred}), with the substitution of $\epsilon_i$  by the 
mean energy 
${\rm \epsilon_{_\odot}\approx 1.35~eV}$ of solar photons, and of ${\rm N_i}$ 
by ${\rm N_{_\odot}}$,
the solar-photon column density along the line of sight. Let
$\theta_{_\odot}$ be the angle between the line of sight and the direction
to the sun. Then:
\begin{equation}
{\rm N_{_\odot}(cos\theta_{_\odot})=
{L_{_\odot}   \over   4\,\pi\,c\,l_{_\odot}\epsilon_{_\odot}}\,  
\left( {\pi-\theta_{_\odot}\over \sin\theta_{_\odot}  } \right) }\, .
\label{sun}
\end{equation}
For a uniform ${\rm cos\,\theta_{_\odot}}$ distribution
during the EGRET data taking, the average column density  is
${\rm \overline N_{_\odot}=\pi 
\,L_{_\odot}/(16\,c\,l_{_\odot}\epsilon_{_\odot})}$, resulting
in a sunlight-induced GBR flux: 
\begin{equation}   
{\rm {dF_\gamma^\odot \over dE}\approx 1.32\times 10^{-4}
         \left[{E\over MeV}\right]~cm^{-2}s^{-1}sr^{-1}MeV^{-1}}.
\label{suntot}
\end{equation}
This contribution is roughly 6\% of our galaxy's
result, Eq.~(\ref{ICSphotons2}).
At ${\rm E_\gamma>
75}$ GeV,  the spectrum of Eq.~(\ref{suntot}) should
steepen, since ICS should then be described by the
Klein--Nishina cross section, and not by its low energy
Thomson limit.

\section{Extragalactic contribution to the GBR}

To estimate this contribution, some concepts and numbers need to be recalled.
Hubble's constan' is ${\rm H_0=100~h~km~s^{-1}Mpc^{-1}}$,
with ${\rm h\sim 0.65}$;
${\rm \Omega_m}$ and $\Omega_\Lambda$ are matter and vacuum
cosmic densities in critical units:
${\rm \Omega\equiv\Omega_m+ \Omega_\Lambda}$;
$y\equiv 1+z$ is the redshift factor.
In a Friedman model, the time to redshift relation is
${\rm dy/dt=-H_0\, f(y)\, y}$, with
${\rm f(y)\equiv [(1-\Omega)\, y^2+\Omega_m\, y^3+\Omega_\Lambda]^{1/2}}$.
The luminosity density of the local universe (Ellis 1997) is
${\rm \rho_{_L}=(2.0\pm 0.4)\times 10^8\, h\, L_{_\odot}\, Mpc^{-3}}$.
The
combination ${\rm \rho_{_L}/L_\star}$ provides an estimate of the average
number density of `Milky-Way-equivalent' galaxies. If the main sources
of CRs are young supernova remnants or gamma-ray bursts,
the CR production rate ought to be proportional (e.g. Wijers et al.~1997)
to the star formation rate ${\rm R_{SFR}[y]}$,
recently measured up to redshift
${\rm z\simeq 4.5}$ (Steidel et al.~1998).

The energy of CMB photons up-scattered by electrons at `epoch y'
is proportional
to ${\rm T(y)=y~T_0}$ and it is subsequently redshifted by the same factor;
hence the spectra from distant galaxies should have
the same energy dependence as from our galaxy. The situation for
SL photons is  more complicated. Young galaxies are bluer than older  
ones, but this effect is overcompensated by the expansion redshift
from a relatively low y, onwards. Yet, at the energies observed by EGRET,
and for the redshift values of $\cal{O}$(1) that  dominate
the extragalactic contribution, all these
blue- and red-shifts simply relocate the photon energy,
while roughly maintaining the slope of the spectrum.
For the sum of all galaxies, we estimate:
\begin{equation}
{\rm {dF_\gamma^{^{EG}}\over dE}\sim
{1\over 4\,\pi}~{dL_\gamma\over E\, dE}~{\rho_{_L}\over L_*}
~{c\over H_{0}}\,
\int_1 {R_{SFR}(y)\over R_{SFR}(0)}~{y\over f(y)}~{dy\over y^3} },
\label{allgals}
\end{equation}
where ${\rm dL_\gamma/dE}$ is to be obtained
from the luminosity of a Milky-Way-like galaxy, Eq.~(\ref{radiation}).
For ${\rm R_{SFR}[y]}$ we interpolate the summary
values of Steidel et al. (1998). In writing Eq.~(\ref{allgals})
we have ignored the fact that, above
 ${\rm E\sim 10}$ GeV,
absorption by $e^+e^-$ production on the IR-to-UV background becomes
relevant (Salamon \& Stecker 1998),
so that the extragalactic contribution should be quenched.

For ${\rm \Omega=\Omega_m=1}$ the value of the
integral in Eq.~(\ref{allgals}) is $\sim 0.82$; it increases to
$\sim 1.08$ for a currently more fashionable universe with
${\rm \Omega=1}$, ${\rm \Omega_\Lambda=0.7}$,
${\rm \Omega_m=0.3}$. For the latter case, the result is:
\begin{equation}
{\rm {dF^{^{EG}}_\gamma\over dE}
= 2.48 \times\! 10^{-4} \left[
 E \over MeV \right]^{-2.10\pm 0.05}
cm^{-2}s^{-1}sr^{-1}MeV^{-1}}\, ,
\label{allgals2}
\end{equation}
roughly $10\%$ of our galaxy's angularly-averaged result, 
Eq.~(\ref{ICSphotons2}).

\section{Detailed comparison with the EGRET data}

Our predictions for the magnitude of the GBR 
and its directional dependence on b and l are shown in 
Figs.~5 and 6. In Fig.~5 we display separately the contributions
from ICS off CMB and SL photons in our galaxy, as well as the uniformly
distributed sunlight and extragalactic components.
 In Fig.~7 we compare 
 the total GBR flux:
\begin{equation}
{\rm {dF_\gamma\over dE}=
{dF_\gamma^0\over dE}+
{dF_\gamma^\star\over dE}+
{dF_\gamma^\odot \over dE}+
{dF_\gamma^{^{EG}}\over dE}
}\, ,
\label{total}
\end{equation}
 obtained by summing Eqs.~(\ref{ICSphotpred}),
Eq.~(\ref{suntot}) and Eq.~(\ref{ICSphotons2}),
with the EGRET data.
Our result is a satisfactory fit to the observed 
magnitude and angular trend of the GBR
 (${\bar\chi^2=0.98}$), a vast improvement over the result for a
constant (extragalactic) ansatz, for which ${\bar\chi^2=2.6}$. 
Although this agreement would be more meaningful, had we used a
more realistic model of starlight,
a more careful treatment may be premature, for the EGRET error
bars are large enough to accommodate considerable variations in
the input modelling. In a previous analysis (Dar et al.~1999), for instance,
we obtained a similarly good fit with an assumed
constant-density, spherical CR-electron halo of radius 25 kpc,
for which the results have the advantage of being simple
analytical functions.

We have neglected various putative extragalactic contributions to
 the GBR. Blazars,
because of their beamed emission, may not be very relevant. 
But CR electrons  injected directly into 
intergalactic space by   active galactic nuclei, 
radio galaxies or gamma ray bursters, may
 give rise to a contribution of comparable magnitude and shape
to that of the CR electrons in external galaxies.
These or other potential sources of GBR photons 
may imply that our parameters ${\rm h_e}$ and ${\rm \rho_e}$
have been overestimated. But this effect cannot be very large,
given our success at describing the non-trivial angular dependence
of the EGRET data.

\section{Conclusions and predictions}

We have presented a simple understanding of the relation
between the spectral indices of cosmic-ray protons, electrons
and the GBR. Accepting the possibility that the CR-electron
distribution in our galaxy may have a scale height larger
than conventionally believed, we have also argued that
the bulk of the GBR could originate in our own galaxy.
Our modelling is extremely simplistic, but quite successful.

The predictions specific to our scenario are:
\begin{itemize}
\item{} The GBR should reflect the
asymmetry of our off-centre position in the Galaxy.
\item{} The halo of Andromeda
should shine in gamma rays above a few MeV, with a luminosity
comparable to that in Eq.~(\ref{radiation}).
Likewise, very nearby star-burst Galaxies, such as M82, 
and radio galaxies with large CR production rates, such as Cygnus A,
 may be visible in gamma rays. 
\item{} If the CR-proton and electron acceleration mechanisms are the same,
the existence of a knee in the observed proton spectrum translates
into a related result for the
power index ${\rm \beta_e}$ of the electron spectrum, which should
steepen above
${\rm E\approx 1.6}$ TeV by $\Delta\beta\sim 1/4$. %
\item{} The GBR  spectrum should not have the sharp cutoff,
above ${\rm E\sim 100}$ GeV,
expected (Salamon \& Stecker 1998) for cosmological sources.
But it should nonetheless steepen around 10--100 GeV, because of
the anticipated ``knee'' in the electron spectrum and of the
energy-dependence of the Klein-Nishina cross section.

\end{itemize}

These features of our scenario should  be testable
when the next generation of cosmic-ray and
$\gamma$-ray satellites (AMS-02 and GLAST) are
operational, hopefully by 2005. In spite of their
maturity, cosmic-ray physics and $\gamma$-ray
astrophysics are still young, and thriving.

ACKNOWLEDGEMENTS\\
We are indebted to G. Bignami, S. Dado and G. Raffelt
for discussions and to I. Moskalenko and A. Strong 
for  permission to reproduce their results in 
our Fig. 5.

\newpage

\noindent
REFERENCES\\
Barwick S. W. et al., 1998, ApJ, 498, 779 \\
Bednarz J., Ostrowski M., 1998, PRL, 80, 3911\\
Berezinskii V. S. et al., 1990, {\it Astrophysics of cosmic rays} (North
Holland, Amsterdam, 1990)\\
Bhattacharjee P., Sigl G., 2000, Phys. Rep., 327, 109\\
Bignami G. et al., 1979, ApJ, 232, 649\\
Bird D. J. et al., 1995, ApJ, 441, 144.\\
Broadbend A., Haslam C. G. T., Osborne, J. L., 1989, MNRAS, 237, 381 \\
Chiang J., Mukerjee R., 1998, ApJ, 496, 772\\
Cohen A., De R\'ujula A., Glashow S. L., 1998, ApJ, 495, 539\\
Connell J. J., 1998,  ApJ, 501, L59\\
Dar A., 1998, astro-ph/9809163, in {\it Proceedings of the Rencontres de la
Vall\'ee d'Aoste,} 1998 (ed. M. Greco), Frascati Physics Series, INFN
Pubs, page 23\\
Dar A., De R\'ujula A., Antoniou N., 1999, astro-ph/9901004\\
Dar A., Plaga R., 1999, A\&A, 349, 259\\
Dixon D. D. et al., 1998, New Astron. 3, 539\\
Duric N., Irwin J., Bloemen H., 1998, A\& A, 331, 428\\ 
Ellis R. S., 1997,  ARA\&A, 35, 389 \\
Evenson P., Meyers P., 1984,  J. Geophys. Res., 89 A5, 2647 \\
Felten J. E., Morrison P., 1966, ApJ, 146, 686\\
Ferrando P. et al., 1996, A\&A 316, 528 \\
Fixsen D.J. et al., 1996, ApJ, 473, 576\\
Gnedin N. Y., Ostriker J. P., 1992, ApJ, 400, 1 \\
Golden R. L. et al., 1984, ApJ, 287, 622\\
Golden R. L., et al., 1994, ApJ, 436, 739 \\
Hartmann D. H., 1995, ApJ, 447, 646\\
Hawking S. W., 1977,  Scientific American, 236, 34\\
Hunter S. D. et al., 1997, ApJ, 481, 205\\
Kazanas D., Protheroe J. P., 1983, Nature, 302, 228\\
Lukasiak  A. et al., 1994, ApJ, 423, 426\\
Mather J. C. et al., 1993,  ApJ, 432, L15\\
Moskalenko I. V., Strong A. W., Reimer O., 1998, astro-ph/9811221\\
Moskalenko I. V., Strong A. W., 2000, ApJ, 528, 357\\
Nishimura  J. et al., 1980, ApJ, 238, 394 \\
Page D. N., Hawking S. W., 1976, ApJ, 206, 1 \\
Pohl M., Esposito J. A., 1998, ApJ, 507, 327\\
Prince T. A., 1979,  ApJ, 227, 676\\
Pritchet C. J., van den Bergh S., 1999, AJ, 118, 833\\
Rudaz  S.,  Stecker  F. W., 1991, ApJ, 368, 40\\
Salamon  M. H.,  Stecker F. W., 1998, ApJ, 493, 547\\
Seo E. S., Ptuskin V. S., 1994, ApJ, 431, 705\\
Silk J., Srednicki M., 1984,  PRL, 53, 264\\
Sreekumar P. et al., 1998, ApJ, 494, 523\\
Stecker F. W., Salamon M. H., 1996, ApJ, 464, 600\\
Stecker F. W., Morgan D. L., Bredekamp J., 1971, PRL, 27, 1469\\
Steidel C.C. et al., 1999, ApJ, 519, 1\\
Strong A., Moskalenko I. V., 1998, ApJ, 509, 212\\
Strong  A. W. et al., 1997, in {\it Proceedings of the 4th Compton 
Symposium},  AIP, 410, 1198\\
Strong A. W., Moskalenko I. V., Reimer O., 1998, astro-ph/9811296\\
Swordy S. P. et al., 1990, ApJ, 330, 625 \\
Takeda M. et al., 1998, PRL, 81, 1163\\
Tang K. K., 1984,  ApJ, 278, 881 \\
Thompson  D. J., Fichtel C. E., 1982, A\&A, 109, 352\\
Webber W. R., 1997, Sp. Sci. Rev., 81, 107 \\
Wiebel-Sooth B., Biermann, P. L., 1998, Landolt-B\"ornstein,
(Springer Verlag, Heidelberg 1998, in press) \\
Wijers R. A. M. J. et al., 1997, MNRAS, 294, L13 \\








\newpage

\newpage
\begin{figure}
\begin{center}
\vspace*{-1.6cm}
\hspace*{-1cm}
\epsfig{file=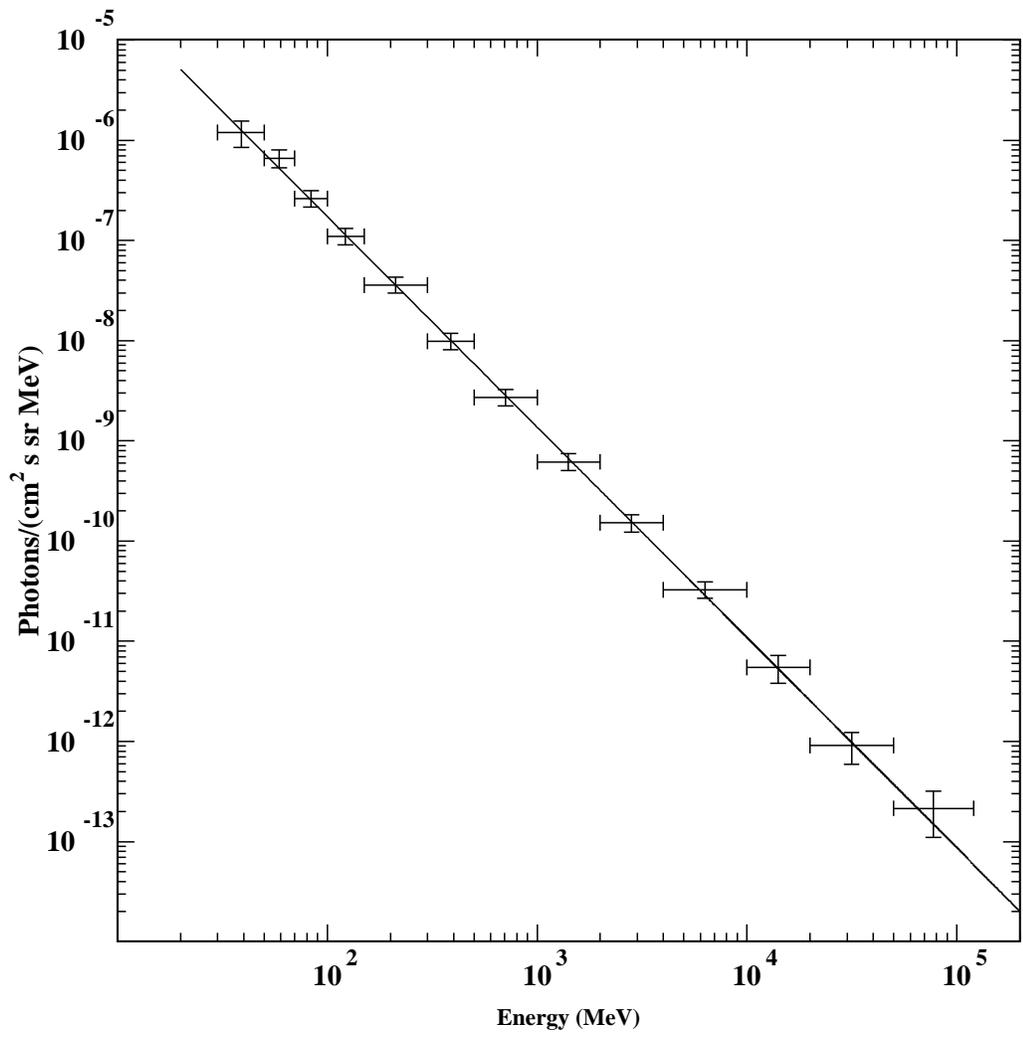,width=15cm}
\caption{Comparison between the spectrum  of
the GBR, measured by EGRET (Sreekumar et al.~1998),
and the prediction for ICS of starlight and the CMB by CR
electrons. The slope is our central prediction, the normalization
is the one obtained for ${\rm h_e}= 20$ kpc, ${\rm \rho_e}= 35$ kpc.}
\vspace*{-0.5cm}
\end{center}
\end{figure}

\begin{figure}
\begin{center}
\vspace*{-1.6cm}
\hspace*{-1cm}
\epsfig{file=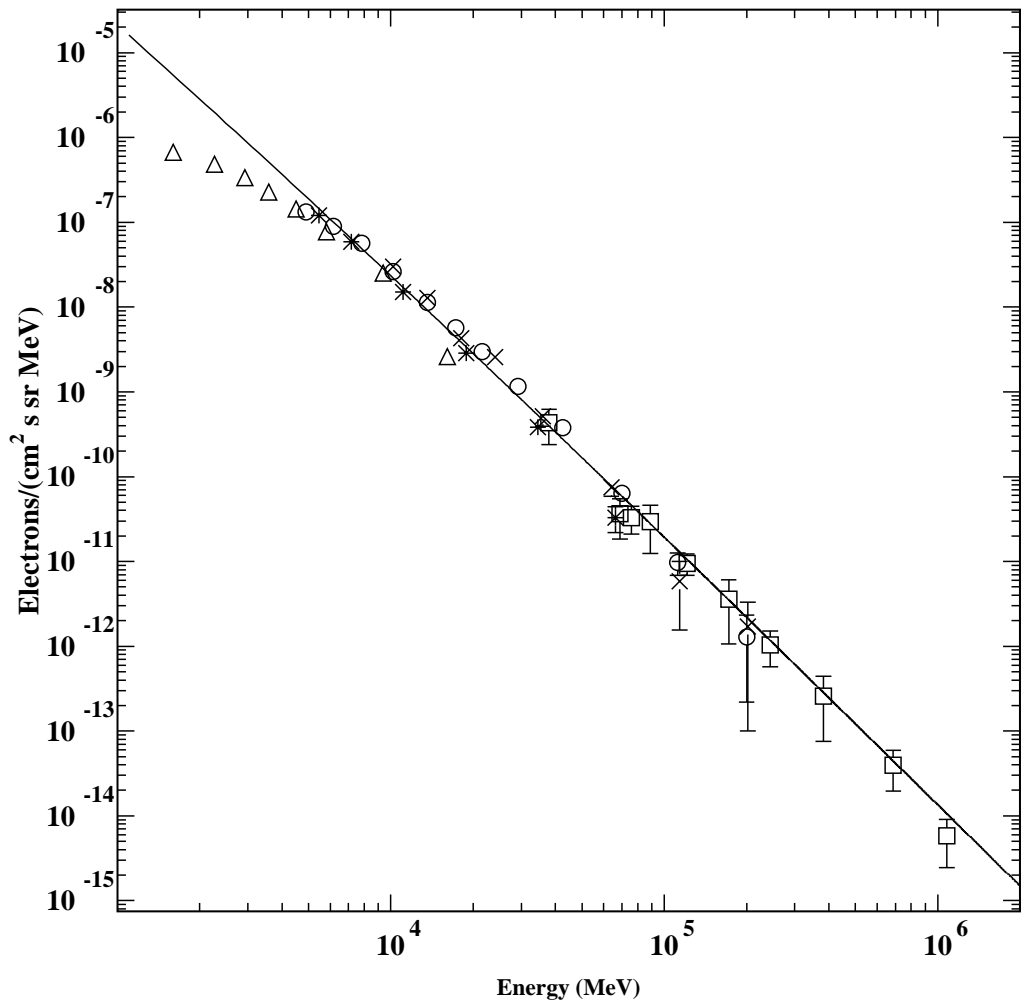,width=15cm}
\caption{The primary cosmic-ray electron spectrum
(Evenson \&
Meyers 1984; Golden et al.~1994; Ferrando et al.~1996) as
measured by Prince 1979 [crosses]; Nishimura et al.~1980 [squares];
Tang 1984 [circles]; Golden et al.~1984 [triangles];  Barwick et
al.~1998 [stars]. The slope is the prediction, the
magnitude is normalized to the data.}
\vspace*{-0.5cm}
\end{center}
\end{figure}

\begin{figure}[t]
\begin{tabular}{cc}
\hskip -0.5truecm
\epsfig{file=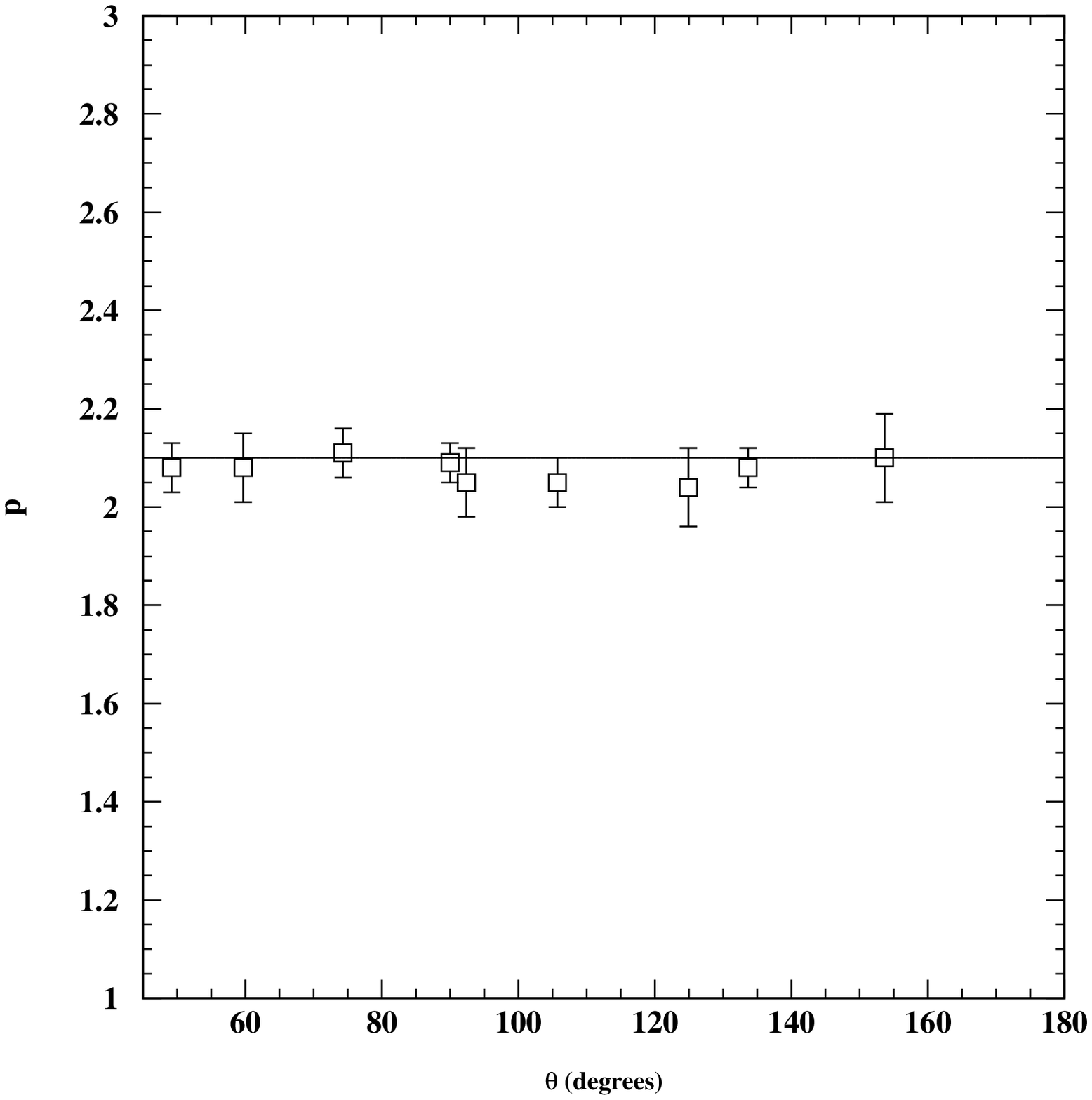,width=8cm} &
\hskip -0.5truecm
\epsfig{file=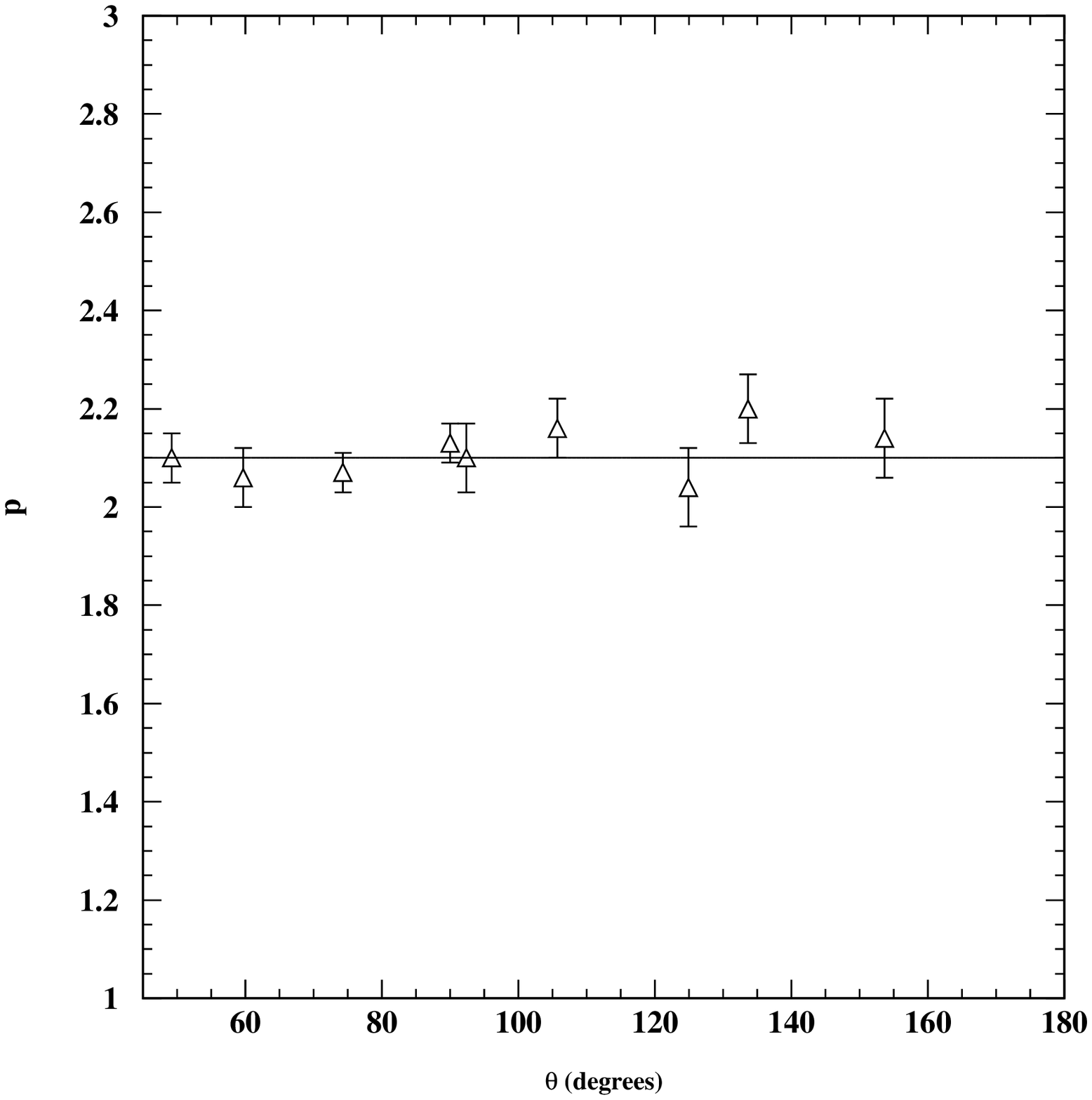,width=8cm}
\vspace{-0.5cm}\\
{\small (a)}            &
\hskip -0.2truecm
{\small (b)} \\
\epsfig{file=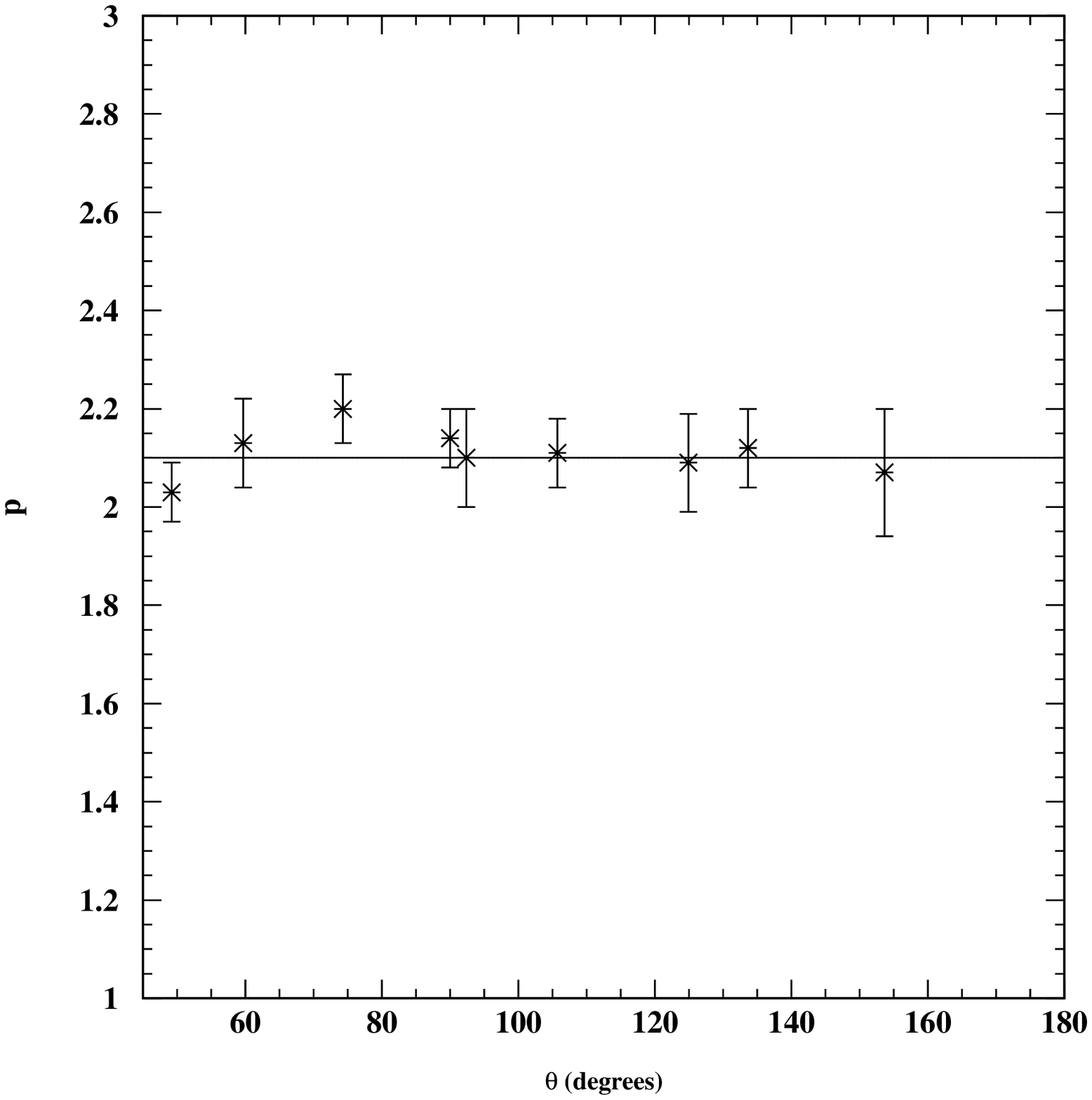,width=8cm} &
\hskip -0.5truecm
\epsfig{file=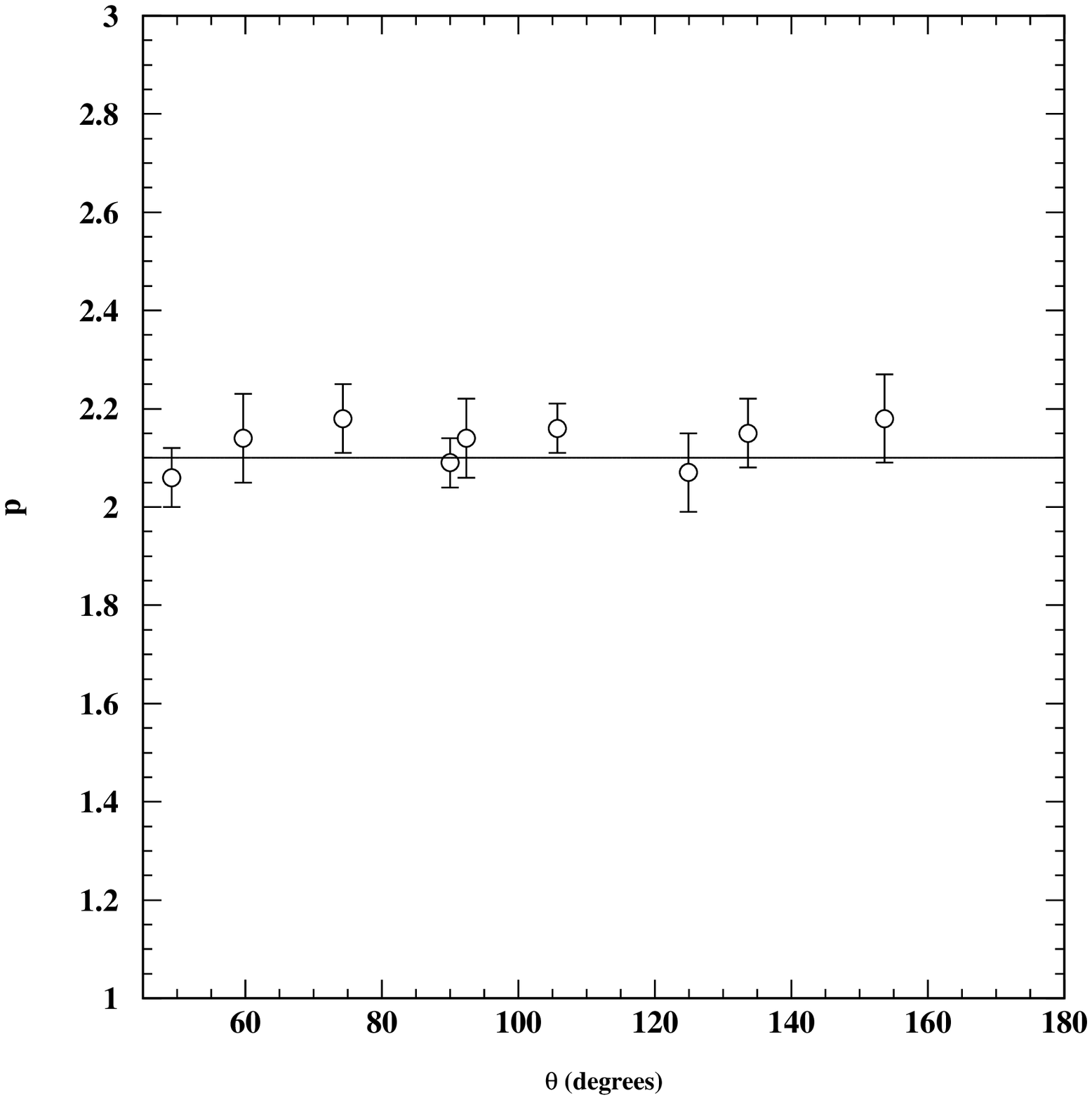,width=8cm}
\vspace{-0.5cm}\\
\hskip 0.2truecm
{\small (c)}            &
\hskip -0.20truecm
{\small (d)}
\end{tabular}
\caption{{
EGRET data on the GBR spectral
index  as a function of  $\theta$
the angle away from the direction of the galactic center.
The line is the predicted spectral index.
The various plots correspond to the individual half-hemispheres.
(a)
${\rm b> 0}$, ${\rm l> 0}$.
(b)
${\rm b> 0}$, ${\rm l< 0}$.
(c)
${\rm b< 0}$, ${\rm l> 0}$.
(d)
${\rm b< 0}$, ${\rm l< 0}$.}}
\label{fig:indexes}
\end{figure}

\begin{figure}[t]
\begin{tabular}{cc}
\hskip -0.5truecm
\epsfig{file=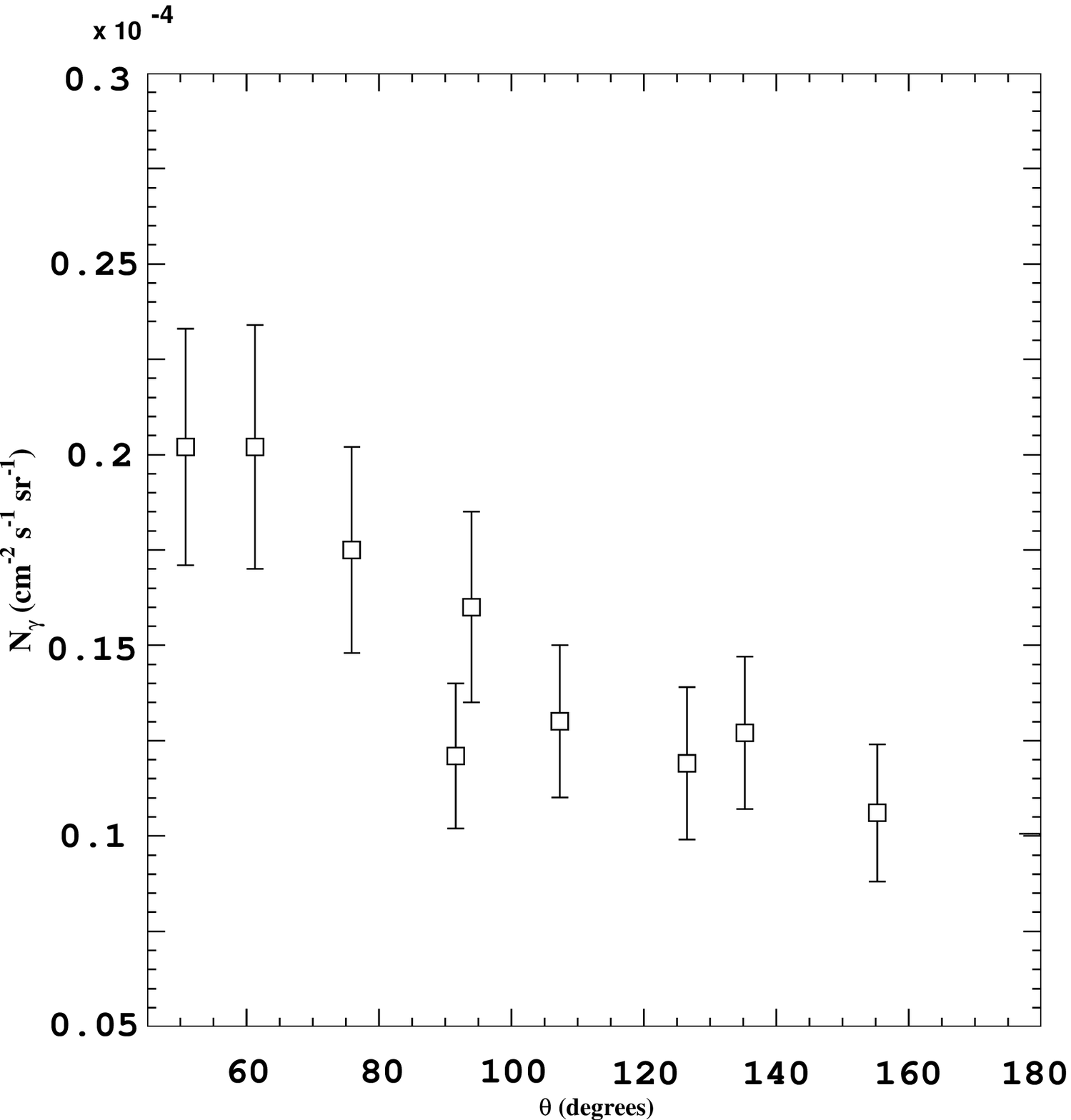,width=7cm} &
\hskip 0.5truecm
\epsfig{file=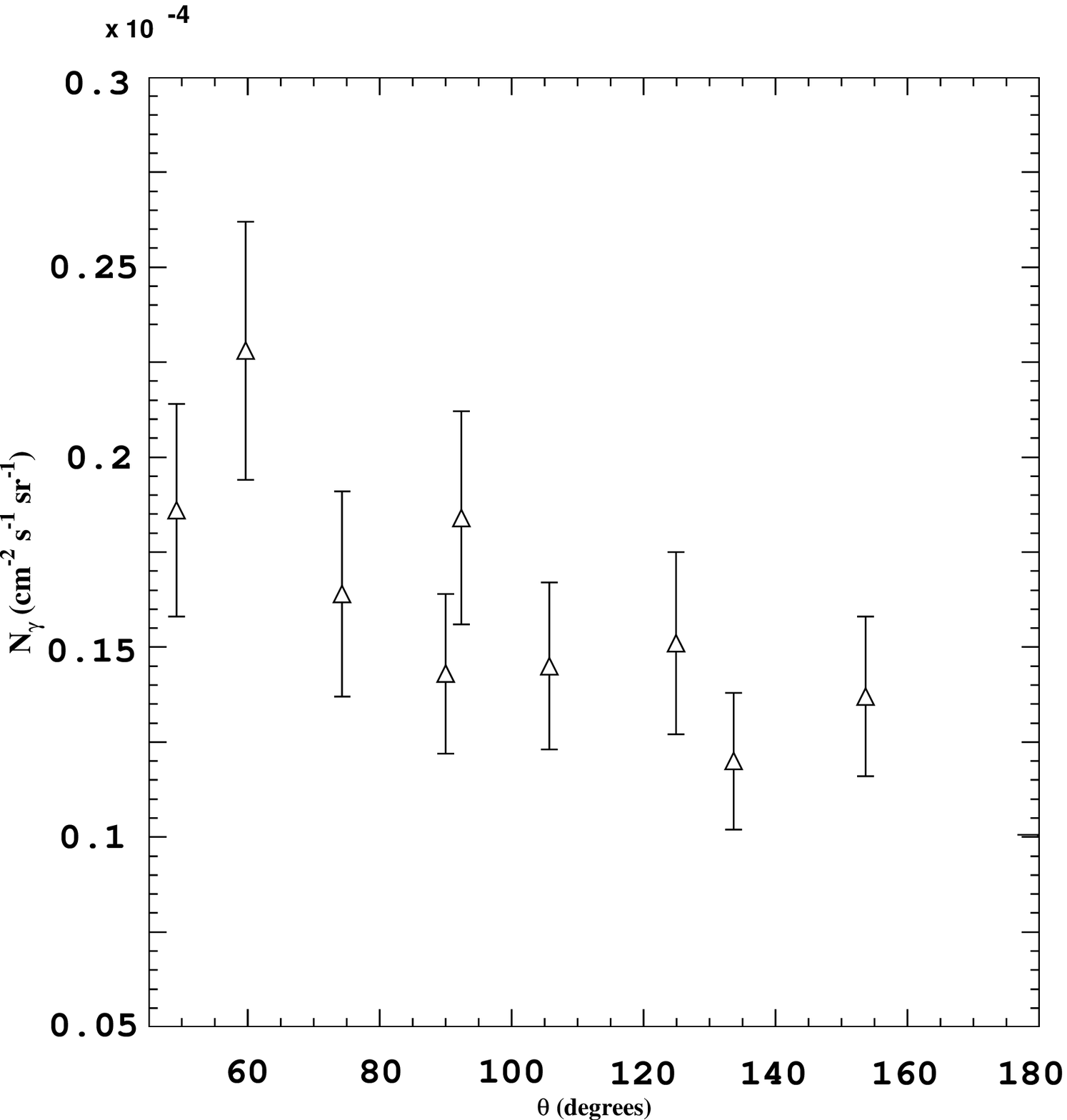,width=7cm}
\vspace{-0.5cm}\\
{\small (a)}            &
\hskip 1.2truecm
{\small (b)} \\
\vspace{-0.9cm}\\
\hskip -0.5truecm
\epsfig{file=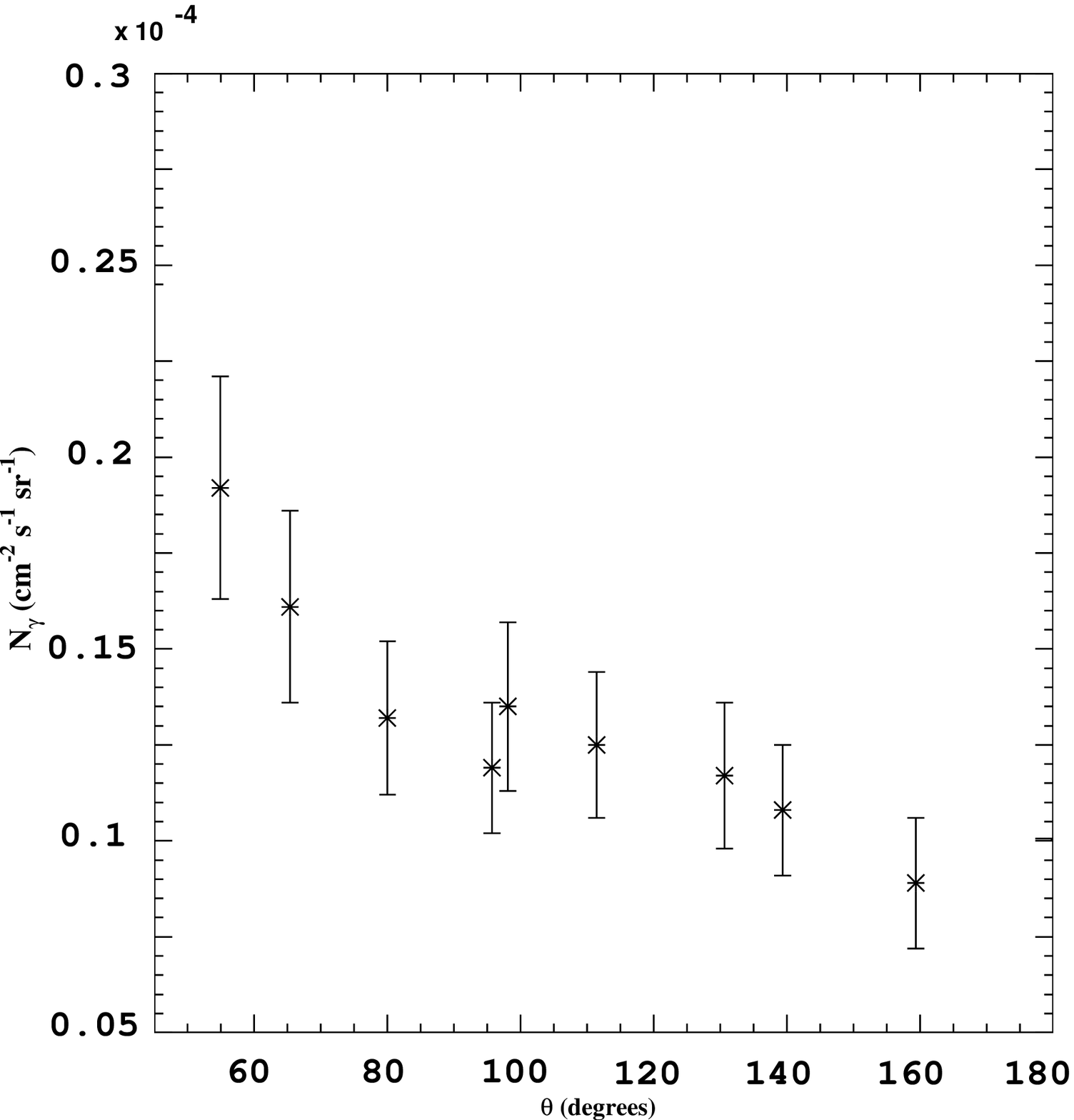,width=7cm} &
\hskip 0.5truecm
\epsfig{file=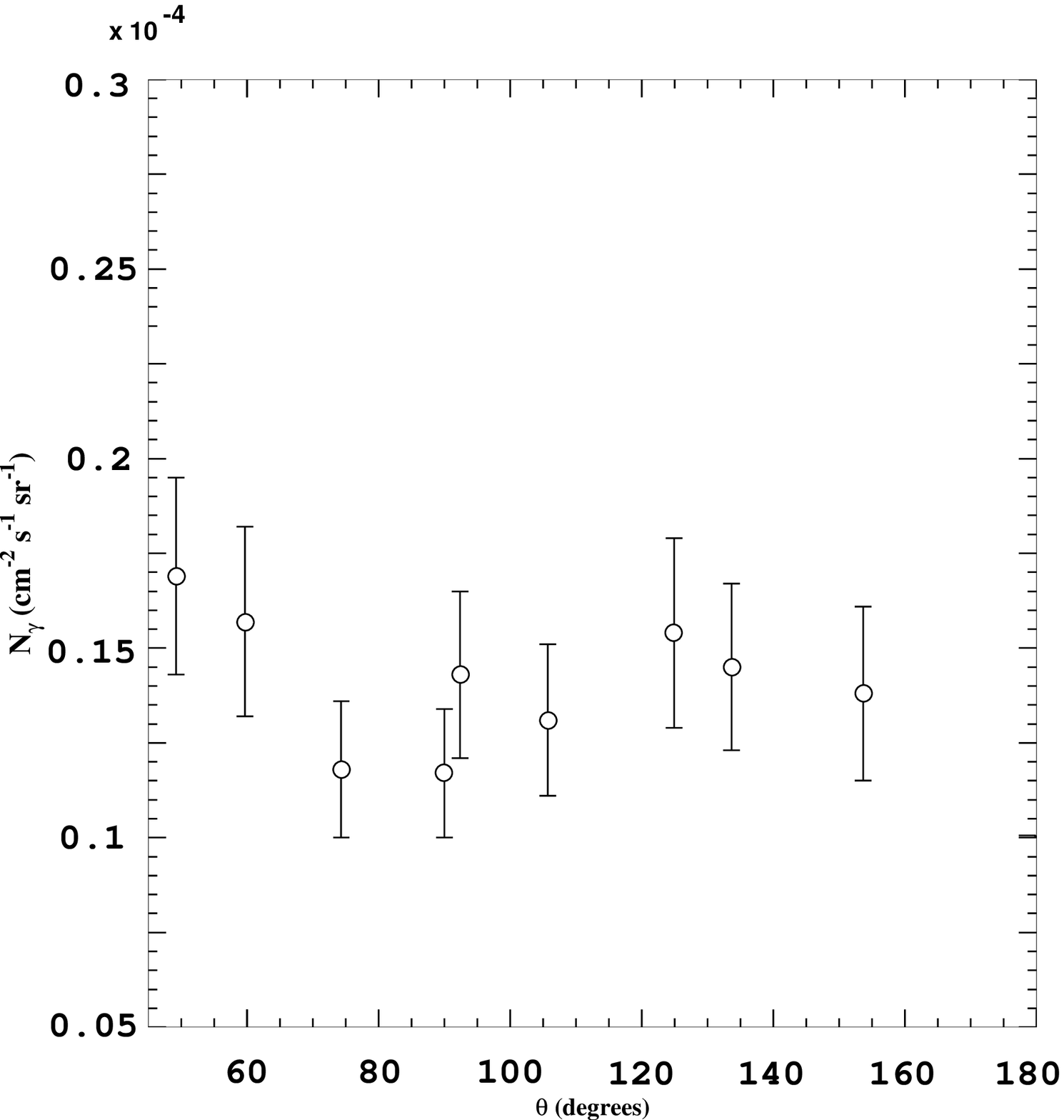,width=7cm}
\vspace{-0.5cm}\\
\hskip 0.2truecm
{\small (c)}            &
\hskip 1.2truecm
{\small (d)}
\end{tabular}
\caption{{EGRET data, organized as in Fig.~3,
for the dependence on $\theta$ of the GBR intensity 
above 100 MeV. }}
\label{fig:intensities1}
\end{figure}

\begin{figure}[t]
\begin{tabular}{cc}
\hskip -0.5truecm
\epsfig{file=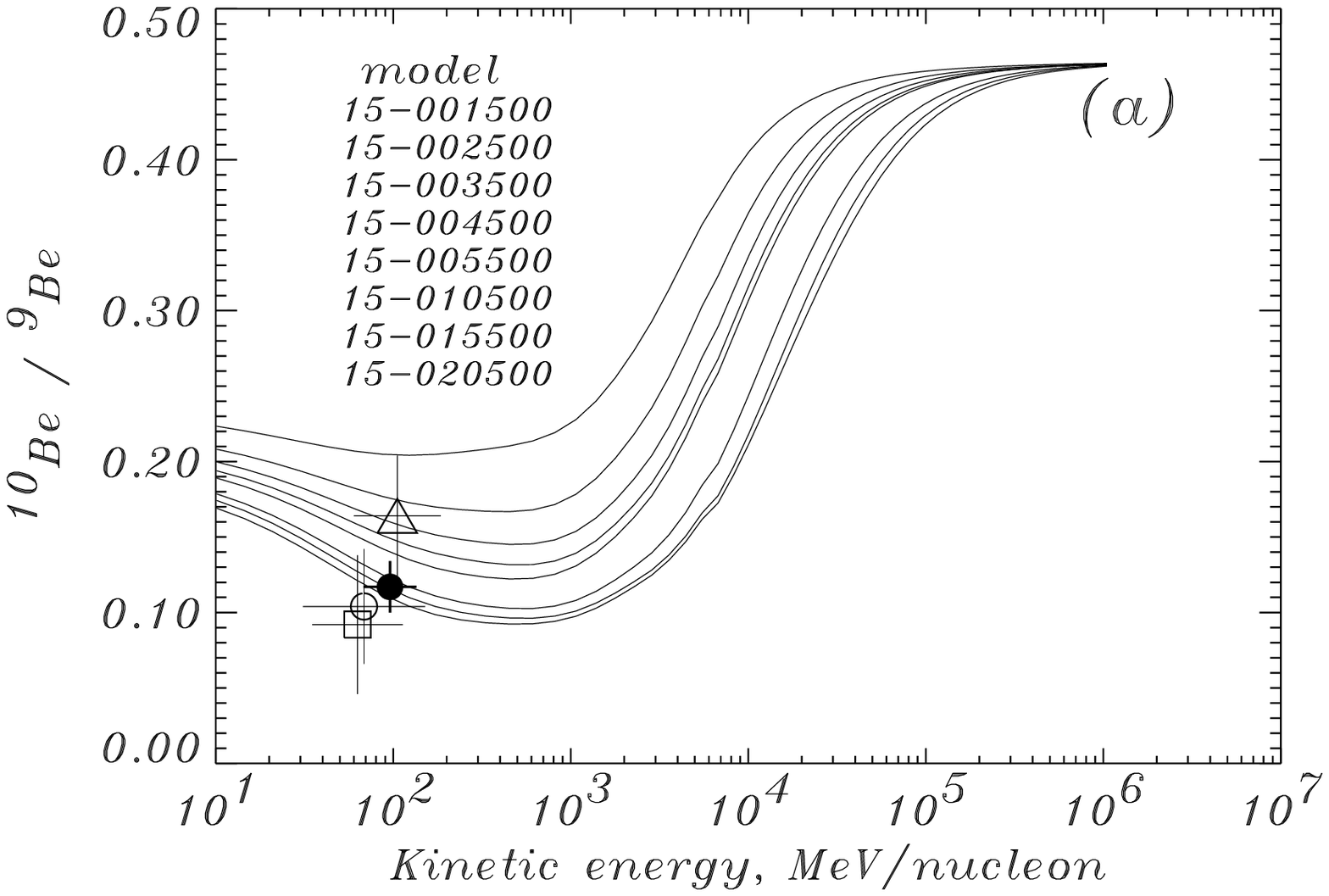,width=8cm} &
\hskip -0.5truecm
\epsfig{file=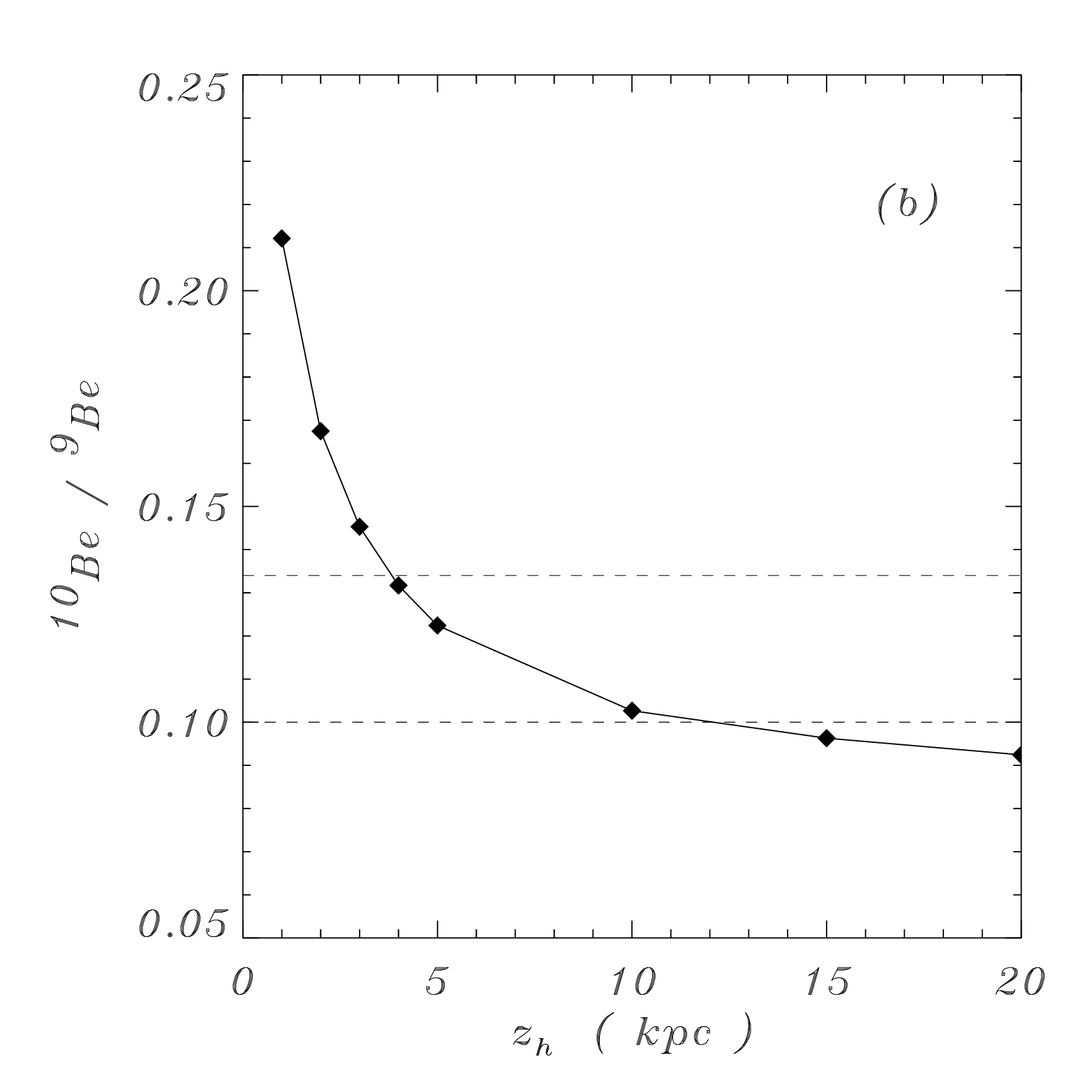,width=8cm}
\vspace{-0.5cm}\\
\end{tabular}
\vspace{0.5 cm}
\caption{{${\rm ^{10}Be/ ^9Be}$ ratio for the diffusive reacceleration models
of Strong and Moskalenko (1998).  (a) As a function of energy for
${\rm z_h=1}$, 2, 3, 4, 5, 10, 15 and 20 kpc. (b) As a function of  
${\rm z_h}$
at 525 MeV/nucleon, the mean interstellar value for the Ulysses data,
whose 1$\sigma$ limits are the dashed lines.
The data points in (a) are from Lukasiak el al.~1994
(square, Voyagers 1,2; open circle, IMP 7/8; triangle, ISEE 3) and Connell
1998 (filled circle, Ulysses).}}
\label{fig:moskastrong}
\end{figure}

\begin{figure}
\begin{center}
\vspace*{-1.6cm}
\hspace*{-1cm}
\epsfig{file=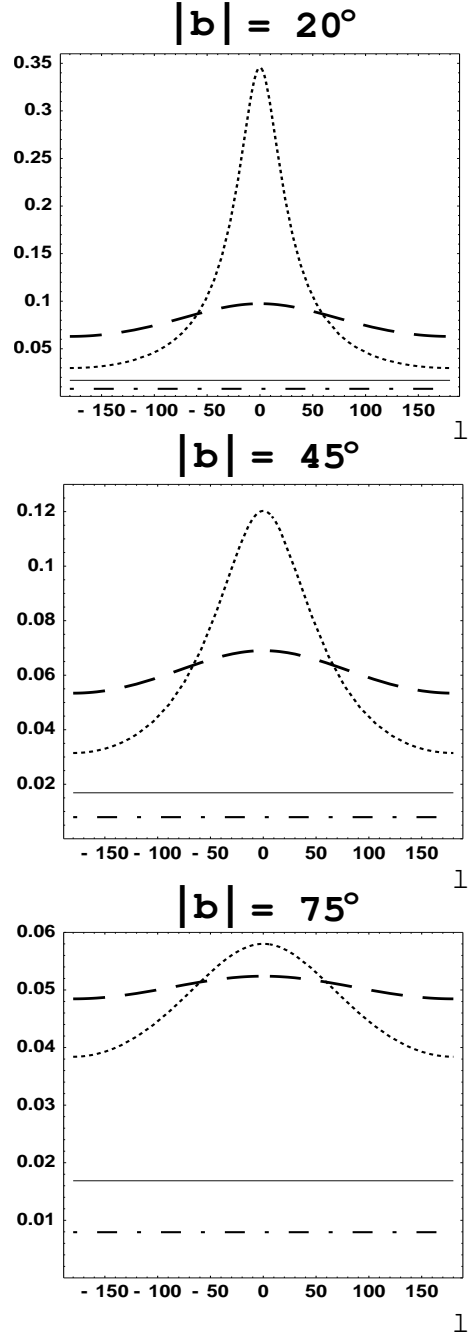,width=6cm}
\caption{Contributions to the GBR flux above 100 MeV
as functions of longitude l, at fixed latitude b, from ICS
of starlight (dotted), and CMB in our galaxy (dashed); from
the total ICS from external galaxies (continuous), and
from sunlight (dot-dashed). The
vertical scale is
$10^4$ times the number of photons/${[\rm cm^2\,s\,sr]}$.
 The results are
for ${\rm h_e=20}$ kpc, ${\rm \rho_e=35}$ kpc.}
\vspace*{-0.5cm}
\end{center}
\end{figure}

\begin{figure}
\begin{center}
\vspace*{-1.6cm}
\hspace*{-1cm}
\epsfig{file=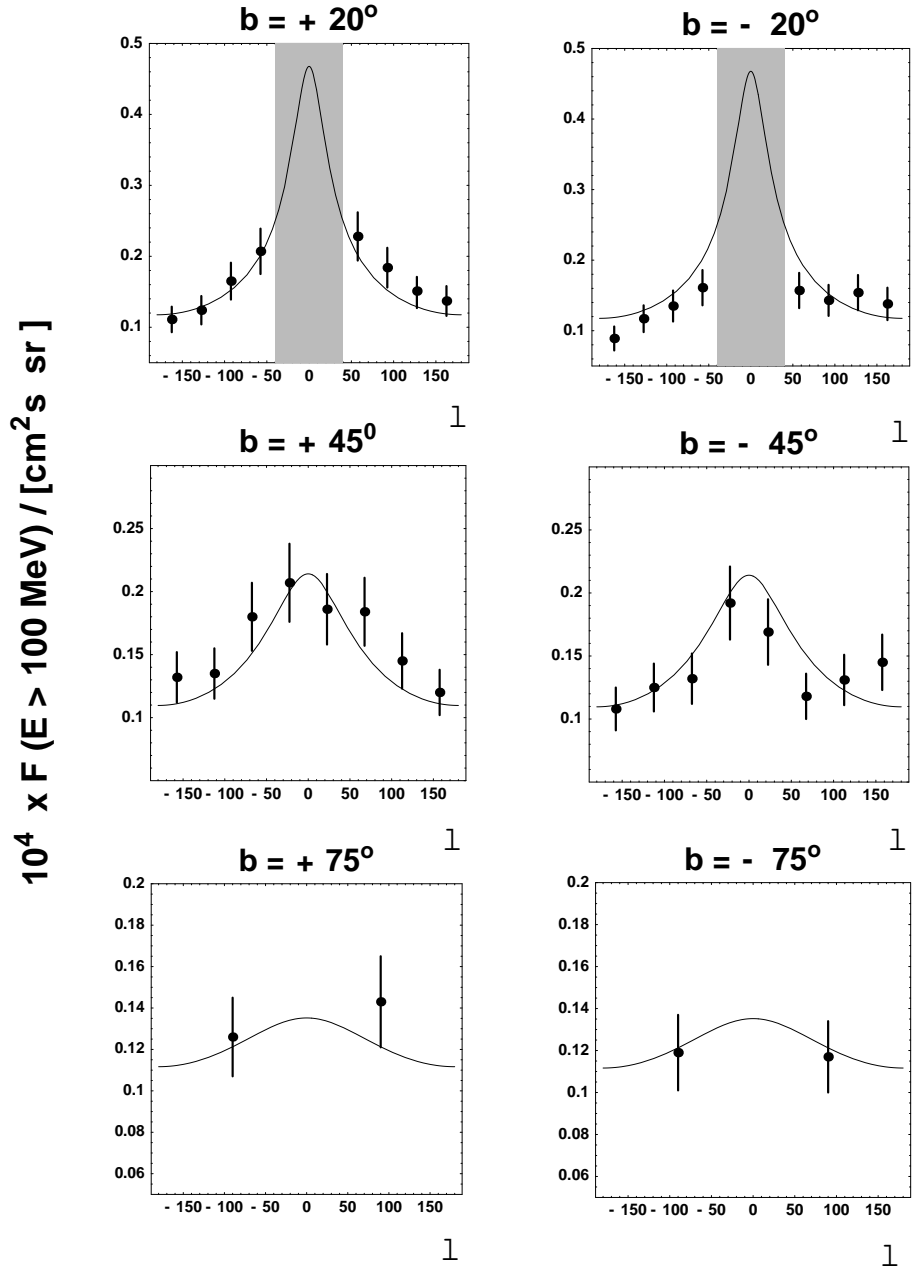,width=12cm}
\caption{The flux of GBR photons above 100 MeV: comparison
between EGRET data and our model 
for ${\rm h_e=20}$ kpc, ${\rm \rho_e=35}$ kpc, as functions
of latitude at various fixed longitudes. The grey domain
is EGRET's mask.}
\vspace*{-0.5cm}
\end{center}
\end{figure}

\end{document}